\documentclass[%
 aip,
 amsmath,amssymb,
 preprint,%
]{revtex4-1}

\usepackage{graphicx}
\usepackage{dcolumn}
\usepackage{bm}

\usepackage[utf8]{inputenc}
\usepackage[T1]{fontenc}
\usepackage{mathptmx}
\usepackage{etoolbox}
\usepackage[hang,nooneline,tight]{subfigure}
\usepackage{tabularx}
\usepackage{array}
\usepackage{booktabs} 
\usepackage{multirow}
\makeatletter
\def\@email#1#2{%
 \endgroup
 \patchcmd{\titleblock@produce}
  {\frontmatter@RRAPformat}
  {\frontmatter@RRAPformat{\produce@RRAP{*#1\href{mailto:#2}{#2}}}\frontmatter@RRAPformat}
  {}{}
}%
\makeatother
\begin{document}

\preprint{}

\title[]{Bayesian optimization for the spanwise oscillation of a gliding flat-plate}
\author{Chunyu Wang}
\author{Zhaoyue Xu}
\author{Xinlei Zhang}
\author{Shizhao Wang}%
 \email{wangsz@lnm.imech.ac.cn}
\affiliation{The State Key Laboratory of Nonlinear Mechanics, Institute of Mechanics, Chinese Academy of Sciences, Beijing 100190, China }%
\affiliation{School of Engineering Sciences, University of Chinese Academy of Sciences, Beijing 101408, China}

\date{\today}

\begin{abstract}
The kinematics of a gliding flat-plate with spanwise oscillation has been optimized to enhance the power efficiency by using Bayesian optimization method, in which the portfolio allocation framework consists of a Gaussian process probabilistic surrogate and a hybrid acquisition strategy. 
We tune three types of acquisition function in the optimization framework and assign three different balance parameters to each acquisition function. 
The design variables are set as the dimensionless oscillating amplitude and reduced frequency of the spanwise oscillation. 
The object function is to maximize the power factor to support a unit weight. 
The optimization results in a maximal power factor of 1.65 when the dimensionless oscillating amplitude and reduced frequency vary from 0 to 1. 
The features of the probabilistic response surface are also examined.
There exists an optimal reduced frequency for the power efficiency at the oscillating amplitudes above 0.40. 
In addition, the higher power efficiency may be obtained by increasing the amplitude beyond 1.00.
\end{abstract}

\maketitle

\section{\label{sec1}Introduction}
There are always great challenges in optimizing a complex black-box system effectively in the aerospace field, such as aerodynamic shape design\cite{mohammadi2004shape,gao2017aerodynamic} and data-driven turbulence modeling\cite{duraisamy2019turbulence,zhang2019progresses}. 
Bayesian optimization methods can be applied to find the best solution to expensive black-box problems  with remarkable advantages\cite{shahriari2015taking}. 
On the basis of a few sampled points, Bayesian optimization visualizes input-output relationships, estimates the optimum as well as its location, and suggests points used to improve the next estimation by an acquisition strategy\cite{jones1998efficient}.
The surrogate model and the acquisition strategy are two main components to integrate the framework of Bayesian optimization. 
 
The foundation of Bayesian optimization is the construction of a surrogate model\cite{forrester2009recent}. The surrogate model utilized is statistical, which means that a probabilistic distribution over objective functions is involved. In other words, the probabilistic model provides the estimation of the unknown object function and the uncertainty of the estimation.
Shahriari et al.\cite{shahriari2015taking} review some probabilistic models in Bayesian optimization.
Above all, the Gaussian process (GP) model, sometimes called kriging, is a widely applied probabilistic model due to its analytic properties, flexibility, and well-calibrated uncertainty\cite{jones2001taxonomy,raissi2017machine}.
In shape design, the GP model works successfully in drag minimization for transonic natural-laminar-flow wings\cite{han2018aerodynamic}, noise minimization for hydrofoil trailing edges\cite{marsden2004suppression,marsden2007trailing}, and power maximization for ducted tail rotors\cite{cao2021numerical}.
In scheme construction, the GP model is introduced to obtain new reconstruction coefficients of the weighted essential non-oscillatory (WENO) method in the finite-difference formulation, which reduces dissipation in smooth regions while preserving high resolution around discontinuities\cite{han2020improved}.  
Furthermore, the GP model can also be transformed for extension to different scenarios. 
On the one hand, it can be extended to multioutput versions modeling several outputs simultaneously and transferring auxiliary information across the outputs\cite{liu2018remarks,lin2021multi}.
On the other hand, the GP model can be extended to multifidelity variants improving the accuracy of prediction with cheap low-fidelity data\cite{lin2021multi}. 

In contrast to the GP model, other probabilistic surrogate models are limited by their intrinsic characteristics.
Random forests\cite{shahriari2015taking} possess scalability and high parallelizability, but they have poor prediction precision at points far from samples. Moreover, they build a discontinuous and nondifferentiable response surface that is detrimental for visualization and gradient-based optimization. 
In addition, deep neural networks\cite{snoek2015scalable} relieve the computational complexity of the GP model with additional effort to design a suitable neural network structure including the number of hidden layers and the number of neurons in each layer.
Thus, the GP model still outperforms the others for simple optimization tasks without too many observations, especially when a smooth response surface is used to investigate the features of the objective function.  

The second key issue in the framework of Bayesian optimization is the acquisition strategy or infill criterion\cite{forrester2009recent}, which forms the active mechanism.
The acquisition function embodying this strategy is cheap to evaluate. Then, suboptimization of the acquisition function can be solved in an efficient way, which drives the outer optimization process. Therefore, it is very important to set up an acquisition strategy for desirable optimization efficiency and quality\cite{ponweiser2008clustered}. 
Generally, the strategy is determined according to two considerations. One is to select the global best point from the current surrogate. Another is to find the point where there is a large amount of uncertainty for local estimation because the evaluation for this underexplored point has more potential to find a better solution and makes the surrogate model more accurate. With the two principles, the acquisition function trades off between exploring global regions and exploiting local regions where expected performance may be found\cite{bradford2018efficient}. Basic acquisition strategies are proposed from the perspective of achieving improvement with respect to the current optimal function value. They are the most likely improvement (MI, not including confidence information), probability of improvement (PI), expected improvement (EI), and upper confidence bound (UCB)\cite{jones1998efficient, jones2001taxonomy,rojas2020survey,zhan2020expected}. 
Emmerich et al.\cite{emmerich2006single} test the above strategies for 10-dimensional generalized Schaffer problems. The results show that the EI strategy yields the best performance but produces the highest computing cost. A test by Liu et al.\cite{liu2012comparison} indicates that the MI strategy performs best in airfoil shape optimization. Different acquisition strategies have different effects in a given application. 

Despite limited experience, the choice of the acquisition strategy is still difficult when a new optimization task appears. A feasible solution avoiding the choice is the hybrid acquisition strategy.  The idea of this strategy is to incorporate several well-established acquisition functions into a portfolio and to select the one with the best adaptability. In this process, multiple suboptimization problems need to be solved, which results in additional computing costs. Nevertheless, the increased cost is negligible compared with that of evaluating the original physical governing equations.  
Hoffman et al.\cite{hoffman2011portfolio} propose a full-information hedging strategy and show its superiority over the individual acquisition strategy through tests for standard functions, sampled functions, and a real physical problem. Vasconcelos et al.\cite{vasconcelos2019no} modify the algorithm by incorporating the memory factor and reward normalization. 

Conclusively, the Bayesian optimization framework combining the GP model with the hybrid acquisition strategy, referred to as the portfolio allocation framework, is a powerful tool in addressing many  missions of interest.
However, to the best of our knowledge, this portfolio allocation framework has rarely been utilized in the research related to fluid dynamics, especially in the kinematic optimization of biological flyers or micro air vehicles (MAVs)~\cite{hassanalian2017classifications}.   

Different optimization methods have been successfully applied to the design of kinematics in biolocomotion.  
Quinn et al.\cite{quinn2015maximizing} use gradient-based method to maximize the propulsive efficiency for a heaving and pitching flexible panel, which gives design criteria of kinematic parameters for the flexible propulsor with the optimal efficiency.  
Ortega-Casanova and Fernandez-Feria\cite{ortega2019maximum} apply the same method to optimize the efficiency for two flapping rigid plates in tandem. It is found that the maximum propulsion efficiency appears when the fore foil has an almost feathering motion and the hind foil fluctuates appropriately through the shedding trailing vortices from the fore foil. 
Zheng et al.\cite{zheng2020multifidelity} adopt a multifidelity GP model and a single acquisition strategy to identify the optimal kinematic parameters for different target aerodynamic forces on a flapping airfoil. Their study reveals the effects of heaving amplitude, flapping frequency, angle of attack amplitude, and stroke angel on the vortex structures and the aerodynamic forces in the two/three-dimensional asymmetrically flapping motions. 

The aim of this work is to extend the application of Bayesian optimation method with the portfolio allocation framework to the design of kinematics of biolocomotion. 
Specifically, we optimize the kinematics of a gliding flat-plate with spanwise oscillation to enhance the aerodynamic efficiency of endurance~\cite{wang2015liftb}.
The method and results are expected to be useful for the design of MAVs. 
The remainder of this paper is organized as follows. 
The portfolio allocation framework of Bayesian optimization is presented in Section 2. 
The optimization framework is tested by an analytical function and then applied to optimize the kinematics of the spanwise oscillating wing in Section 3.
Finally, conclusions are drawn in Section 4.

\section{\label{sec2} Bayesian optimization framework}

\subsection{\label{sec2.1} Gaussian processes}
Bayesian optimization is a surrogate-based optimization method. Because of the excellent nonlinear fitting ability, a GP is utilized to generate the target regression. As a nonparametric model, ${GP}\left( {{\mu}_{0}},k \right)$ is constructed from the prior mean function ${{\mu}_{0}}:{\chi} \mapsto \mathbb{R}$ and the positive-semidefinite covariance function $k:{\chi}\times{\chi} \mapsto \mathbb{R}$ \cite{shahriari2015taking}. The former represents a possible offset of the expectation of the objective function, while the latter determines the smoothness and amplitude. In this work, we adopt the smooth squared exponential covariance function:

\begin{equation}
    k\left( \bm{\xi},\bm{\xi}' \right)={\sigma _{f}}^{2}\exp \left( -\frac{1}{2{l^2}}{{\left\| \bm{\xi }-\bm{\xi}' \right\|}^{2}} \right),
\end{equation}
where ${\sigma _{f}}^{2}$ is the maximum covariance and $l$ is the length parameter that controls the effect of the distance between the two design vectors $\bm{\xi}$ and $\bm{\xi}'$. 

According to the nature of the GP, any finite subset of random variables obeys the multivariate normal distribution. Here, the observations to the training and prediction points constitute this subset. Additionally, Gaussian noise is considered in each observation. Consequently, the basic assumptions for the GP model can be written as

\begin{equation}
    \mathbf{f}\left| {{\bm{\xi}}_{1:n}},{\bm{\xi}^{*}}_{1:m} \right.\sim {\mathcal N}\left( \mathbf{m},\mathbf{K} \right)
    \label{assumption1},
\end{equation}

\begin{equation}
    \mathbf{q}\left| \mathbf{f},\bm{\varepsilon} \right.\sim {\mathcal N}\left( \mathbf{f},{{\sigma }^{2}}\mathbf{I} \right)
    \label{assumption2},
\end{equation}
where ${\bm{\xi}}_{i}$ is the known design vector (training point) and ${{\bm{\xi}}^{*}}_{i}$ is the unknown design vector (prediction point). Here, $\mathbf{f}$ and $\mathbf{q}$ are vectors consisting of unknown object values and noisy observation values on $n$ training points ${\bm{\xi}}_{1:n}$ and $m$ prediction points ${{\bm{\xi}}^{*}}_{1:m}$, respectively; $\mathbf{m}$ is the prior mean vector with elements ${{m}_{i}}={{\mu}_{0}}\left( {\hat{\bm{\xi}}}_{i} \right)$, where ${\hat{\bm{\xi}}}_{i}$ is a training point or a prediction point; $\mathbf{K}$ is the prior covariance matrix with elements ${{K}_{i,j}}=k\left( {{\hat{\bm{\xi}}}_{i}},{{\hat{\bm{\xi}}}_{j}} \right)$; and $\bm{\varepsilon}$ is the noise vector with its elements ${\varepsilon}_{i}$ following the Gaussian distribution ${\mathcal N}\left( 0,{{\sigma}^{2}} \right)$.
For ease of expression, some items in the relations~(\ref{assumption1}) and (\ref{assumption2}) are expanded into partitioned forms, as shown below:

\begin{equation}
    \begin{aligned}
        \mathbf{f}=\begin{bmatrix}{{f}_{1:n}} \\ {f^{*}}_{1:m} \end{bmatrix}= \begin{bmatrix} f\left( {{\bm{\xi}}_{1:n}} \right) \\ f\left( {{\bm{\xi}}^{*}}_{1:m} \right) \end{bmatrix} \in {{\mathbb{R}}^{n+m}},
        \;
        \mathbf{q}= \begin{bmatrix}{{q}_{1:n}} \\ {q^{*}}_{1:m} \end{bmatrix} \in {{\mathbb{R}}^{n+m}},
        \\
        \mathbf{m}= \begin{bmatrix}{{m}_{1:n}} \\ {m^{*}}_{1:m} \end{bmatrix} = \begin{bmatrix} {{\mu}_{0}}\left( {\bm{\xi}}_{1:n} \right) \\ {{\mu}_{0}}\left( {{\bm{\xi}}^{*}}_{1:m} \right) \end{bmatrix} \in {{\mathbb{R}}^{n+m}},
        \;
        \bm{\varepsilon}= \begin{bmatrix} {{\varepsilon}_{1:n}} \\ {{\varepsilon}^{*}}_{1:m}
        \end{bmatrix} \in {{\mathbb{R}}^{n+m}}
        \label{fourvectors},
    \end{aligned}
\end{equation}

\begin{equation}
    \mathbf{K}=\begin{bmatrix}{{K}_{n\times n}} & {K^{*}}_{n\times m} \\ {K^{*}}_{m\times n} & {K^{**}}_{m\times m} \end{bmatrix} \in {{\mathbb{R}}^{\left( n+m \right)\times \left( n+m \right)}}
    \label{covariancematrix},
\end{equation}
where $\mathbf{f}$, $\mathbf{q}$, $\mathbf{m}$ and $\bm{\varepsilon}$ are decomposed into the training and prediction subvectors and $\mathbf{K}$ is divided into three components of ${K}_{n\times n}$, ${K^{*}}_{n\times m}$ (${K^{*}}_{m\times n}$ is the transpose of ${K^{*}}_{n\times m}$), and ${K^{**}}_{m\times m}$ interpreting the correlation of all pairs of points in ${\left\{ \left( {{\bm{\xi}}_{i}},{{\bm{\xi}}_{j}} \right) \right\}}_{i,j=1:n}$, ${\left\{ \left( {{\bm{\xi}}_{i}},{{{\bm{\xi}}^{*}}_{j}} \right) \right\}}_{i=1:n,\;j=1:m}$, and ${\left\{ \left( {{{\bm{\xi}}^{*}}_{i}},{{{\bm{\xi}}^{*}}_{j}} \right) \right\}}_{i,j=1:m}$, respectively.

Combining the relations~(\ref{assumption1}) and (\ref{assumption2}) together with substituting Equations~(\ref{fourvectors}) and (\ref{covariancematrix}), we can obtain

\begin{equation}
    \begin{aligned}
        \begin{bmatrix} {{q}_{1:n}}  \\ {q^{*}}_{1:m} \end{bmatrix} 
        \left| {{\bm{\xi}}_{1:n}},{{\bm{\xi}}^{*}}_{1:m} \right.
        =\left(  \begin{bmatrix} {{f}_{1:n}}  \\ {{f^{*}}_{1:m}} \end{bmatrix} +
        \begin{bmatrix} {{\varepsilon}_{1:n}}  \\ {{{\varepsilon}^{*}}_{1:m}} \end{bmatrix} \right)
        \left| {{\bm{\xi}}_{1:n}},{{\bm{\xi}}^{*}}_{1:m} \right.
        \\
        \sim {\mathcal N}
        \left( \begin{bmatrix} {{m}_{1:n}}  \\ {{m^{*}}_{1:m}} \end{bmatrix} ,
        \begin{bmatrix} {{K}_{n\times n}}+{{\sigma}^{2}}{{I}_{n}} & {K^{*}}_{n\times m}  \\
        {K^{*}}_{m\times n} & {K^{**}}_{m\times m}+{{\sigma}^{2}}{{I}_{m}} \end{bmatrix} \right).
    \end{aligned}
\end{equation}

Then, we apply the property of conditioning Gaussians, resulting in the following distribution:

\begin{equation}
    {q^{*}}_{1:m} \left| {{q}_{1:n}},{{\bm{\xi}}_{1:n}},{{\bm{\xi}}^{*}}_{1:m} \right.
    \sim {\mathcal N}
    \left( {{\mu}^{*}}_{1:m},{{\Sigma}^{*}}_{m\times m} \right),
\end{equation}
where ${{\mu}^{*}}_{1:m}$ is the posterior mean vector and ${{\Sigma}^{*}}_{m\times m}$ is the posterior covariance matrix. They are expressed as

\begin{equation}
    {{\mu}^{*}}_{1:m}=
    {m^{*}}_{1:m} + {K^{*}}_{m\times n} {{\left( {{K}_{n\times n}}+{{\sigma}^{2}}{{I}_{n}} \right)}^{-1}} \left( {{q}_{1:n}}-{{m}_{1:n}} \right)
    \label{posteriormean},
\end{equation}

\begin{equation}
    {{\Sigma}^{*}}_{m\times m}=
    \left( {K^{**}}_{m\times m}+{{\sigma}^{2}}{{I}_{m}} \right) - {K^{*}}_{m\times n}
    {{\left( {{K}_{n\times n}}+{{\sigma}^{2}}{{I}_{n}} \right)}^{-1}} {K^{*}}_{n\times m}
    \label{posteriorvariance}.
\end{equation}

Once the training set ${{\mathcal D}_{1:n}}={{\left\{ \left( {{\bm{\xi}}_{i}},{{q}_{i}} \right) \right\}}_{i=1:n}}$ is created, the best estimate for ${q^{*}}_{1:m}$ and its uncertainty are calculated through equations~(\ref{posteriormean}) and (\ref{posteriorvariance}), respectively, which is the modeling of the GP. In particular, when only one prediction point is considered $\left( m=1 \right)$, we can derive the posterior mean function ${\mu} \left( \bm{\xi} \right)$ and variance function ${{\sigma}^{2}}\left( \bm{\xi} \right)$ (or the standard deviation function ${\sigma} \left( \bm{\xi} \right)$).

\subsection{\label{sec2.2} Acquisition strategy}

After the probabilistic surrogate model is built, we can use the statistical information provided by the model to create the acquisition function, and maximizing this function can help search for the promising optimal point. The suboptimization problem can be formulated as
\begin{equation}
    {{\bm{\xi}}_{n+1}}=\arg {{\max}_{\bm{\xi}\in {\chi}}\alpha \left( \bm{\xi};{{\mathcal D}_{1:n}} \right)}.
\end{equation}

As the active mechanism of Bayesian optimization, the acquisition function makes use of the posterior prediction function to find the candidate point following a certain criterion. The acquisition function has various forms according to different criteria. Here, we employ three confidence information-assisted strategies, including the PI, EI, and UCB, as described above. For an arbitrary design vector $\bm{\xi}$, these acquisition functions are expressed as follows:
\begin{equation}
    \begin{aligned}
        {{\alpha}_{PI}} \left( {\bm{\xi}};\;{{\mathcal D}_{1:n}} \right)
        & =
        prob\left( f\left( \bm{\xi} \right) \ge {\tau} +{{\zeta}_{PI}} \right)
        ={\Phi} \left( \frac{ {\mu} \left( \bm{\xi} \right) - {\tau} - {{\zeta}_{PI}} }
        { {\sigma} \left( \bm{\xi} \right) } \right)
        ,
        \\
        {{\alpha}_{EI}} \left( {\bm{\xi}};\;{{\mathcal D}_{1:n}} \right)
        & =
        \left\{ \begin{matrix}
        \left( {\mu} \left( \bm{\xi} \right) - {\tau} - {{\zeta}_{EI}} \right) \cdot 
        {\Phi} \left( \frac{ {\mu} \left( \bm{\xi} \right) - {\tau} - {{\zeta}_{EI}} } 
        { {\sigma} \left( \bm{\xi} \right) } \right)
        +
        {\sigma} \left( \bm{\xi} \right) \cdot 
        {\phi} \left( \frac{ {\mu} \left( \bm{\xi} \right) - {\tau} - {{\zeta}_{EI}} }
        { {\sigma} \left( \bm{\xi} \right) } \right) 
        & {\sigma} \left( \bm{\xi} \right) > 0  \\
        0 
        & {\sigma} \left( \bm{\xi} \right) = 0   \end{matrix} \right. 
        ,
        \\
        {{\alpha}_{UCB}} \left( {\bm{\xi}};\;{{\mathcal D}_{1:n}} \right)
        & =
        {\mu} \left( \bm{\xi} \right) + {{\zeta}_{UCB}} \cdot {\sigma} \left( \bm{\xi} \right)
        \label{acqs},
    \end{aligned}
\end{equation}
where $\tau$ is the incumbent optimal target, ${\Phi} \left( \centerdot \right)$ is the standard normal cumulative distribution function, and ${\phi} \left( \centerdot \right)$ is the standard normal probability density function. In addition, ${\zeta}_{PI}$, ${\zeta}_{EI}$, and ${\zeta}_{UCB}$ denote the balance parameters used to trade off between global exploration and local exploitation.

Different acquisition criteria have different adaptability to a model with specific spatial characteristics. Moreover, the preferred strategy may change with the advancement of sequential optimization. Therefore, compared with using the single constant acquisition function, a better alternative is to dynamically pick a superior function from the prescribed portfolio. The robust strategy is leveraged in this study.
We define an acquisition function portfolio that contains the PI, EI, and UCB. Each type of acquisition function is assigned three different balance parameters, as shown in Table~\ref{table1}.

\begin{table}
    \caption{\label{table1}Acquisition function portfolio.}
    \begin{tabular}{c c c}
        \hline
        \hline
        \quad\quad Name \quad\quad\quad&\quad\quad\quad Type \quad\quad\quad\,&Balance parameter\\
        \hline
        Acq 1&PI&0.00\\
        Acq 2&PI&0.01\\
        Acq 3&PI&0.10\\
        Acq 4&EI&0.00\\
        Acq 5&EI&0.01\\
        Acq 6&EI&0.10\\
        Acq 7&UCB&1.00\\
        Acq 8&UCB&1.50\\
        Acq 9&UCB&2.00\\
        \hline
        \hline
    \end{tabular}
\end{table}

\subsection{\label{sec2.3} Framework integration}

In Bayesian optimization, the probabilistic surrogate model offers the distribution of the target for every point in the design space, and the selected acquisition functions use these statistics to guide the search for the best point. The GP-Hedge algorithm\cite{hoffman2011portfolio} and its modification\cite{vasconcelos2019no} adopt the acquisition function portfolio strategy to improve robustness.

\begin{figure}[!htb]
    \centering
    \includegraphics[width=.9\textwidth]{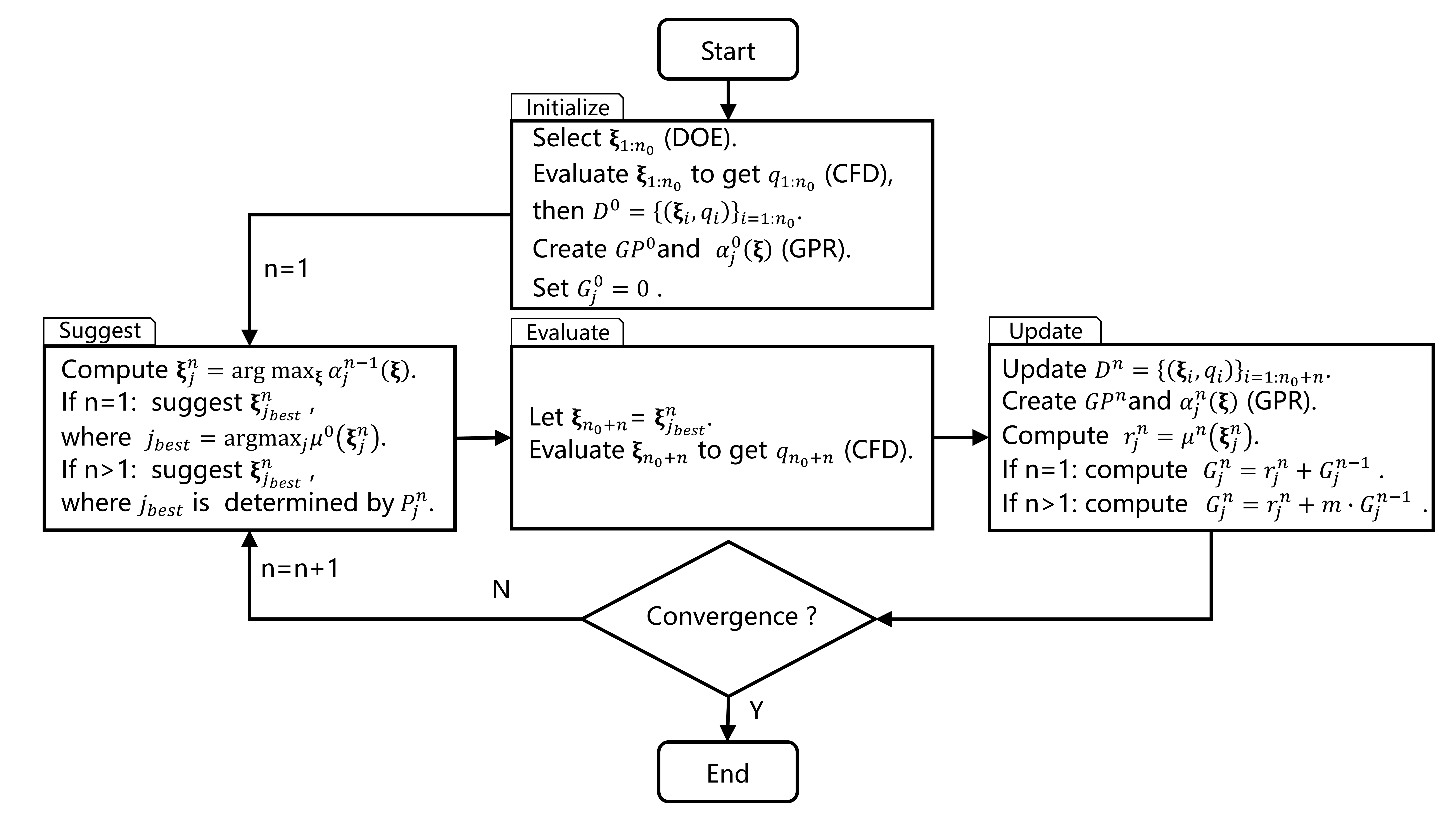}
    \caption{\label{BO_framework} Portfolio allocation framework for Bayesian optimization.}
\end{figure}

The frame structure of Bayesian optimization is detailed in Figure~\ref{BO_framework}, where the superscript $n$ represents the iteration index ($n=0$ means objects generated in the initialization) and the subscript $j$ represents the acquisition function index ($j=1,2,\cdots ,9$). To find a reasonable acquisition function in the beginning and ensure sufficient optimization efficiency, we incorporate the coarse sampling based on the design of experiments (DOE)\cite{yondo2018review} into the initialization. The technique used here is the optimal Latin hypercube method, which ensures that the sampling points are evenly distributed in the design space as much as possible. For the design vector with two variables, that is, ${\bm{\xi}}_{i}={{\left[ \begin{matrix}{{A}_{i}} & {{k}_{i}} \end{matrix} \right]}^{T}}$, 20 points are sampled by the DOE technique and evaluated by the CFD method, resulting in the initial training set ${{\mathcal D}^{0}}={{\left\{ \left( {{\bm{\xi}}_{i}},{{q}_{i}} \right) \right\}}_{i=1:{{n}_{0}}}}$ $\left( {{n}_{0}}=20 \right)$. Then, we create the initial surrogate model ${GP}^{0}$ (its posterior mean function and covariance function are ${{\mu}^{0}}\left( \bm{\xi} \right)$ and ${{\sigma}^{0}}\left( \bm{\xi} \right)$, respectively) and the corresponding acquisition functions $\alpha _{j}^{0}\left( \bm{\xi } \right)$ by Gaussian process regression (GPR). As in the original algorithm, the initial gains $G_{j}^{0}$ are set to 0. 

The aforementioned initialization is followed by a loop composed of `suggest', `evaluate', and `update' steps. The first step is used to query the potential point ${{\bm{\xi}}_{{j}_{best}}}^{n}$ from the candidates ${{\bm{\xi}}_{j}}^{n}$ ($j=1,2,\cdots ,9$) according to the specified criterion, the second step is used to evaluate the selected point ${{\bm{\xi}}_{{j}_{best}}}^{n}$(${\bm{\xi}}_{{n_0}+n}$) through the numerical simulation (CFD) method for obtaining its target value ${q}_{{n_0}+n}$, and the third step is used to sequentially update the data set ${\mathcal D}^{n}$, the surrogate model $GP^n$, the acquisition functions ${\alpha_{j}}^{n}$, and the gains ${G_{j}}^{n}$. It needs to be emphasized that differences exist between iterations $n=1$ and $n>1$. In the `suggest' step, the criterion to select ${\bm{\xi }_{{j}_{best}}}^{n}$ when $n=1$ is maximizing the initial posterior mean function ${{\mu}^{0}}\left( \bm{\xi} \right)$, while the point ${\bm{\xi}_{{j}_{best}}}^{n}$ can be found with the probability ${P_j}^{n}={\exp \left( \eta \cdot {g_j}^{n-1} \right)}/{\sum\nolimits_{{j}'=1}^{9}{\exp \left( \eta \cdot {g_{j}'}^{n-1} \right)}}$, where hyperparameter $\eta =4.0$ and normalized gains ${g_j}^{n-1}={\left[ {G_j}^{n-1}-{{\max }_{j}}\left( {G_j}^{n-1} \right) \right]}/{\left[ {{\max }_{j}}\left( {G_j}^{n-1} \right)-{{\min }_{j}}\left( {G_j}^{n-1} \right) \right]}$, when $n>1$. In the `update' step, the gains ${G_j}^{n}$ are equal to the current rewards ${r_j}^{n}={{\mu}^{n}}\left( {\bm{\xi}_{j}}^{n} \right)$ for $n=1$, while ${G_j}^{n}={r_j}^{n}+m\cdot {G_j}^{n-1}$ for $n>1$. Here, the memory factor $m$ of 0.70 is used to reduce the effect of the past rewards. In this study, the loop stops when the given number of iterations is reached.

\section{\label{sec3}Application to kinematic optimization}

In practice, there is limited knowledge on the new modeling system to be optimized, particularly the nature of the response space within the system. The response space has diverse properties in terms of the modality, plateau region, valley region, separability, and dimensionality\cite{jamil2013literature}. The treatment of complexity has become a trend of method development. Here, the involvement of hybrid acquisition provides an approach to improve the robustness of the Bayesian optimization method. The application to kinematic optimization of the portfolio allocation framework is carried out after the test on a benchmark function. 

\subsection{\label{sec3.1}Test on the cosine mixture function}

To examine the optimization framework in Section 2, a benchmark problem is investigated. Note that the initialization process not only accelerates the search for the optimal solution but also provides some a priori information regarding the distribution of the target in the design space. Beneficial information is used to select an appropriate benchmark function. That is, a test function with comparable optimization complexity can be applied to verify the optimization framework. In this line of thought, we select the cosine mixture function
\begin{equation}
    f \left( \bm{\xi} \right) =
    0.1 \left[ \cos \left( 5 \pi {{\xi}_{1}} \right) + \cos \left( 5 \pi {{\xi}_{2}} \right) \right]
    - \left( {\xi_1}^{2} + {\xi_2}^{2} \right),
\end{equation}
\begin{figure}[!htb]
    \centering
    \subfigcapskip=-10pt
    \subfigure[\;]{
    \label{TestFunction}
    \includegraphics[width=.51\textwidth]{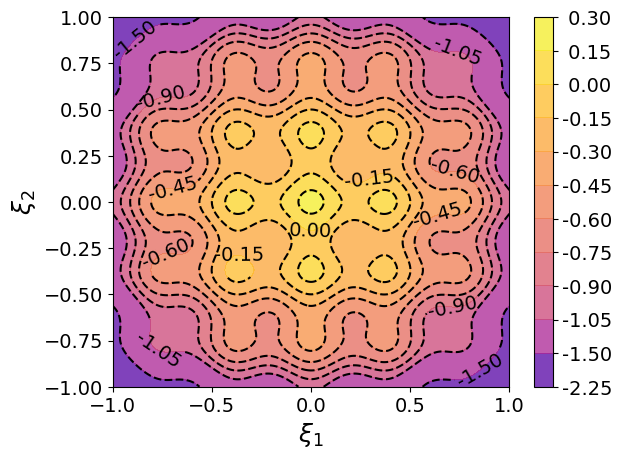}}
    \quad
    \subfigure[\;]{
    \label{TestResidual}
    \includegraphics[width=.42\textwidth]{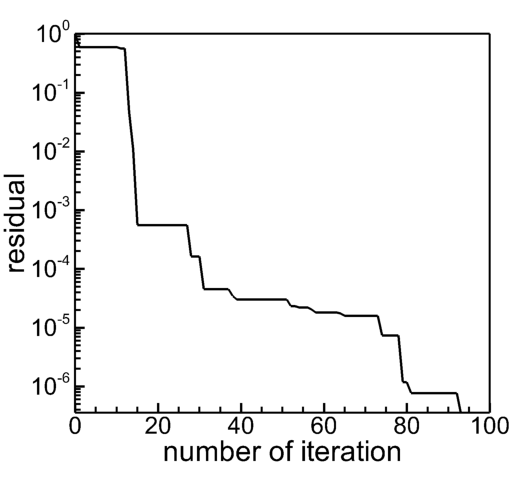}}
    \caption{\label{Test} Validation of the Bayesian optimization framework. (a) Contour of the cosine mixture function used to test the framework. (b) Convergence history of test function optimization.}
\end{figure}
where input variables ${{\xi}_{1}},{{\xi}_{2}}\in \left[ -1,1 \right]$, as shown in Figure~\ref{TestFunction}. The global maximum is 0.20 when ${{\xi}_{1}}$ and ${{\xi}_{2}}$ are both 0. We use the framework to optimize this test function. The convergence history shown in Figure~\ref{TestResidual} verifies the validity of the method. Furthermore, when the optimization loop iterates only 15 times, the incumbent maximum has a value of 0.199836. Therefore, the maximum number of iterations is set to 15, which is considered sufficient in the current study.

\subsection{\label{sec3.2}Efficiency maximization for spanwise oscillation}
Inspired by biological flight and confirmed by numerical simulation, the compound motion of superimposing the spanwise oscillation on the flapping motion can significantly increase the lift by adjusting the vortex system\cite{wang2015liftb}. To effectively use this flow control method, the nonlinear interaction between the spanwise motion and flow structures needs to be further explored, especially its impact on power efficiency. In this paper, the portfolio allocation framework is used to design the spanwise oscillation of a translating plate. 

\subsubsection{\label{sec3.2.1}Design variable selection}
We consider a simplified geometric model in typical flight conditions as follows. A flat rectangular plate with an aspect ratio (AR=2) moves forward at a constant speed $U$ and angle of attack (AoA=$25^{\circ}$). At the same time, different spanwise oscillations characterized by oscillation amplitude and frequency are imposed on the plate. The two kinematic parameters quantify the intensity of oscillation, which has a significant influence on aerodynamic performance. Following the definition of spanwise motion in Wang et al.\cite{wang2015liftb}, the current work assumes that the center of the plate changes its position with time in harmonic form as
\begin{equation}
    y\left( t \right)=A\sin \left( 2kt \right)
    \label{oscillation},  
\end{equation}
where $y$ is the coordinate in the spanwise direction, $t$ is the time, $A$ is the oscillation amplitude, and $k$ is the reduced frequency (the four variables are dimensionless). 
We define a design space $\chi$ with two design variables of $A$ and $k$. Each design variable is set between 0 and 1 for the preliminary study.

\subsubsection{\label{sec3.2.2}Unsteady aerodynamic performance evaluation}

The evaluation of the unsteady aerodynamic target function is based on the incompressible Navier-Stokes equations in the formulation of the immersed boundary method. The reference length and velocity are the chord length $c$ and the uniform oncoming flow velocity $U$, respectively. We adopt the Reynolds number ($Re=300$) consistent with that of the MAV flight. The immersed boundary method is implemented by the discrete stream function and the parallel computing strategy\cite{wang2011immersed,wang2013parallel}.
The numerical scheme and mesh are reported at great length in previous work\cite{wang2011immersed,wang2015liftb}. Through the CFD method, we can obtain the time history of the drag coefficient ($C_D(t)$), the lift coefficient ($C_L(t)$), and the side-force coefficient ($C_S(t)$). 

We consider the efficiency of endurance under spanwise oscillation here. An applicable criterion is created with the constraint of the load balance\cite{wang2008aerodynamic}. 
The derived power factor is the objective function measuring the efficiency. Its expression is
\begin{equation}
    PF=\frac{{\overline{{C_L}\left( t \right)}}^{1.5}}{{\overline{{C_D}\left( t \right)}}+ \frac{1}{T}\int_{0}^{T}{{C_S}\left( t \right) 2 A k \left(-\cos \left( 2kt \right) \right) dt} }
    \label{powerfactor},
\end{equation}
where the overbar denotes the period-averaged force coefficient and $T$ is the dimensionless oscillating period. 

\subsubsection{\label{sec3.2.3}Optimization procedure implementation}

\begin{figure}[!htb]
    \centering
    \includegraphics[width=.45\textwidth]{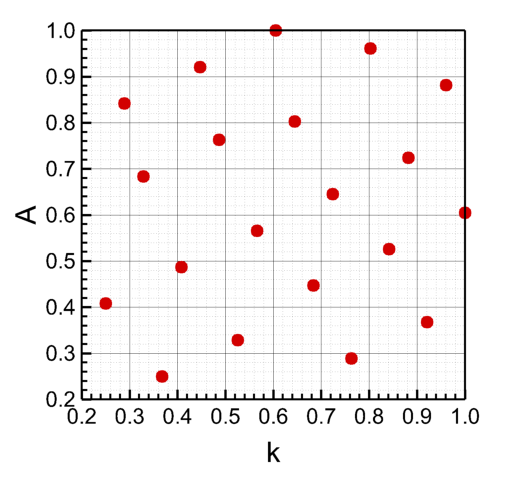}
    \caption{\label{LHD_A05k05} Initial sampling in the design space before the optimization iterations.}
\end{figure}

In the portfolio allocation framework, the first step is to execute the DOE. 
As displayed in Figure~\ref{LHD_A05k05}, the sampled points are equally distributed throughout the design space owing to the optimal Latin hypercube technique. Every sampled point corresponds to a spanwise oscillating configuration. 
Next, we can use the CFD method to evaluate the aerodynamic performances for every configuration. It can be observed in Table~\ref{table2} that the power factor ranging from 1.00 to 2.00 has a significant difference when the oscillating parameters change. There is sufficient space to enhance the efficiency by designing kinematics. 

\begin{table}[!htb]
    \caption{\label{table2} Aerodynamic performances of the initial sampled points.}
    \begin{tabular}{c c c c c c}
        \hline
        \hline
        \quad\quad Case\quad\quad\quad & \quad\quad$A$\quad\quad\quad & \quad\quad$k$\quad\quad\quad & \quad\quad${\overline{{C_D}\left( t \right)}}$ \quad\quad\quad & \quad\quad${\overline{{C_L}\left( t \right)}}$ \quad\quad\quad & \quad\quad$PF$\quad\quad\quad\\
        \hline
        1    & 0.84 & 0.29 & 0.60     & 0.90     & 1.38 \\
        2    & 1.00 & 0.61 & 0.79     & 1.36     & 1.62 \\
        3    & 0.68 & 0.33 & 0.60     & 0.90     & 1.38 \\
        4    & 0.65 & 0.72 & 0.70     & 1.14     & 1.43 \\
        5    & 0.88 & 0.96 & 0.83     & 1.47     & 1.24 \\
        6    & 0.29 & 0.76 & 0.52     & 0.70     & 1.04 \\
        7    & 0.37 & 0.92 & 0.57     & 0.80     & 1.06 \\
        8    & 0.33 & 0.53 & 0.52     & 0.70     & 1.08 \\
        9    & 0.92 & 0.45 & 0.71     & 1.17     & 1.62 \\
        10   & 0.25 & 0.37 & 0.50     & 0.66     & 1.07 \\
        11   & 0.45 & 0.68 & 0.59     & 0.87     & 1.22 \\
        12   & 0.61 & 1.00 & 0.72     & 1.15     & 1.18 \\
        13   & 0.57 & 0.57 & 0.65     & 1.01     & 1.44 \\
        14   & 0.41 & 0.25 & 0.51     & 0.68     & 1.09 \\
        15   & 0.80 & 0.65 & 0.75     & 1.28     & 1.59 \\
        16   & 0.76 & 0.49 & 0.69     & 1.13     & 1.59 \\
        17   & 0.49 & 0.41 & 0.59     & 0.87     & 1.33 \\
        18   & 0.96 & 0.80 & 0.84     & 1.48     & 1.44 \\
        19   & 0.72 & 0.88 & 0.75     & 1.26     & 1.32 \\
        20   & 0.53 & 0.84 & 0.66     & 1.01     & 1.25 \\
        \hline
        \hline
    \end{tabular}
\end{table}

\begin{figure}[h]
    \centering
    \subfigcapskip=-10pt
    \subfigbottomskip=0pt
    \subfigure[\;]{
    \label{BO_PFm_0}
    \includegraphics[width=.45\textwidth]{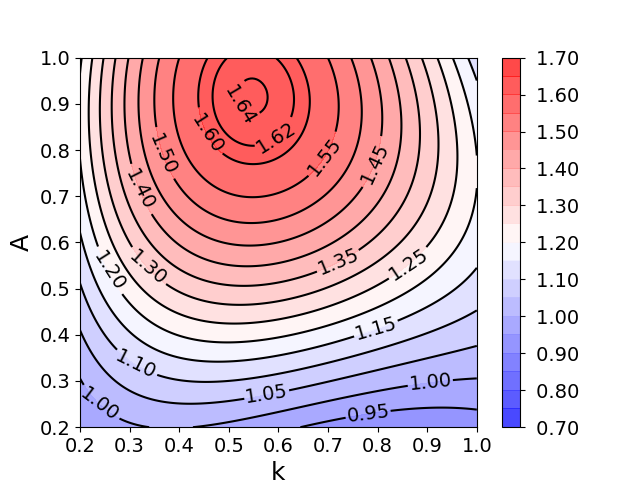}}
    \subfigure[\;]{
    \label{BO_PFm_5}
    \includegraphics[width=.45\textwidth]{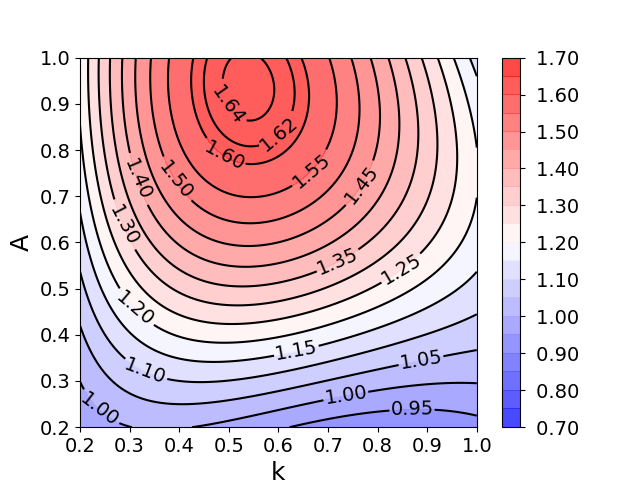}}
    \subfigure[\;]{
    \label{BO_PFm_10}
    \includegraphics[width=.45\textwidth]{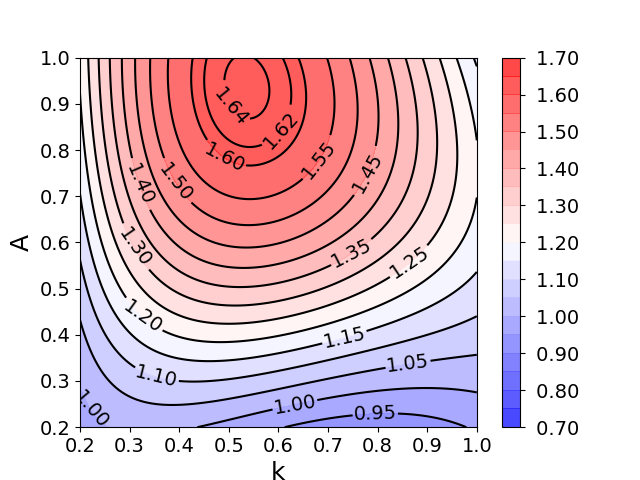}}
    \subfigure[\;]{
    \label{BO_PFm_15}
    \includegraphics[width=.45\textwidth]{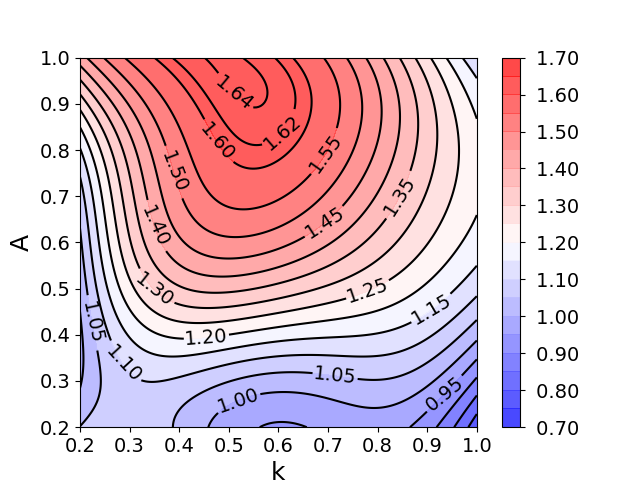}}
    \caption{\label{BO_PFm} Distribution of the power factor in the spanwise oscillating parameter space ($k, A$) as the optimization process advances. (a) Initialization. (b) The $5_{th}$ iteration. (c) The $10_{th}$ iteration. (d) The $15_{th}$ iteration.}
\end{figure}

The optimization iteration begins after the initialization. The iteration process advances with two processes: searching for the optimum and refining the model. Figure~\ref{BO_PFm} shows the estimation for the power factor at different stages, including the initialization and three iterations. 
The initial response surface in Figure~\ref{BO_PFm_0} provides a coarse observation for the true distribution of the power factor. There is a peak in the high-amplitude and medium-frequency region. 
The additive points suggested by the acquisition strategy are clustered around the peak. In the iteration process, the position of the peak transfers gradually to the boundary of the amplitude $A=1.00$.  
Ultimately, the optimization framework locates the maximal power factor of 1.65 at the point where $A=1.00$ and $k=0.49$. Due to the smoothness of the fitted response surface, we can observe some distribution features of the power factor. When the oscillating amplitude is larger than 0.40, a single local optimum appears as the reduced frequency increases. When the reduced frequency is fixed, the power factor under the configurations with $A>1.00$ may be larger than that under the configurations with  $A<1.00$.     

\begin{figure}[h]
    \centering
    \subfigcapskip=-10pt
    \subfigbottomskip=0pt
    \subfigure[\;]{
    \label{BO_PFs_0}
    \includegraphics[width=.45\textwidth]{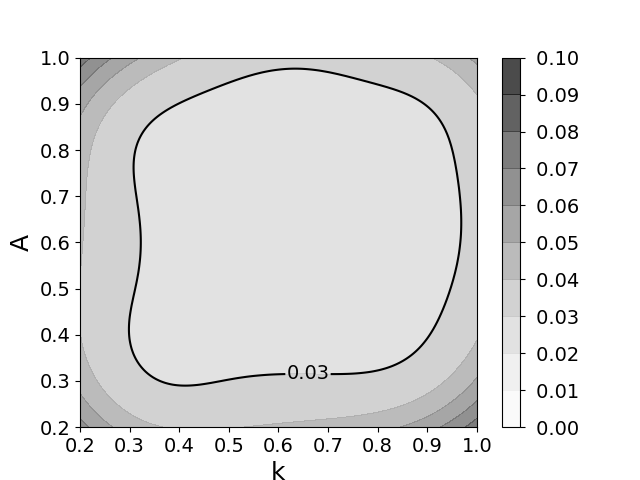}}
    \subfigure[\;]{
    \label{BO_PFs_5}
    \includegraphics[width=.45\textwidth]{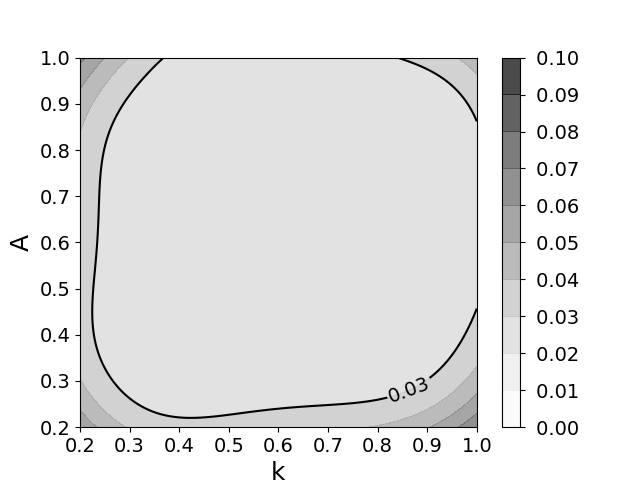}}
    \subfigure[\;]{
    \label{BO_PFs_10}
    \includegraphics[width=.45\textwidth]{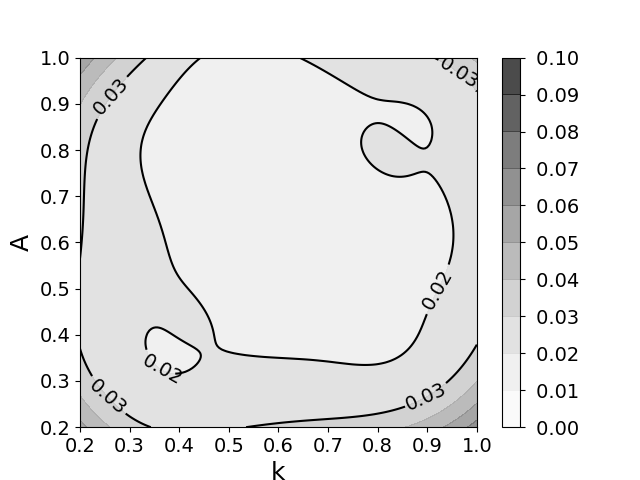}}
    \subfigure[\;]{
    \label{BO_PFs_15}
    \includegraphics[width=.45\textwidth]{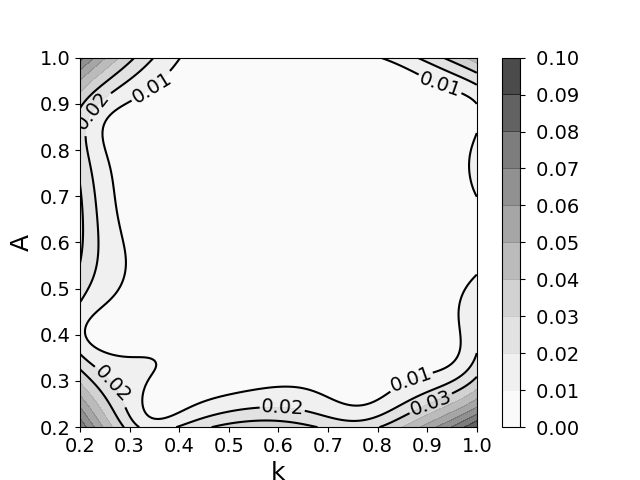}}
    \caption{\label{BO_PFs} Uncertainty of the power factor in the spanwise oscillating parameter space ($k, A$) as the optimization process advances. (a) Initialization. (b) The $5_{th}$ iteration. (c) The $10_{th}$ iteration. (d) The $15_{th}$ iteration.}
\end{figure}

In addition, the estimation uncertainty of the power factor is exhibited in Figure~\ref{BO_PFs}. The standard deviation interpreting the uncertainty decreases in the investigated region with the update of the GP model. This indicates that the accuracy of the model improves with the involvement of additional points. The more accurate regression model would allow analysis of the high-power-efficiency mechanism, which is beyond the scope of this work.    

\begin{figure}[h]
    \centering
    \subfigcapskip=-10pt
    \subfigure[\;]{
    \label{Figure19a}
    \includegraphics[width=.45\textwidth]{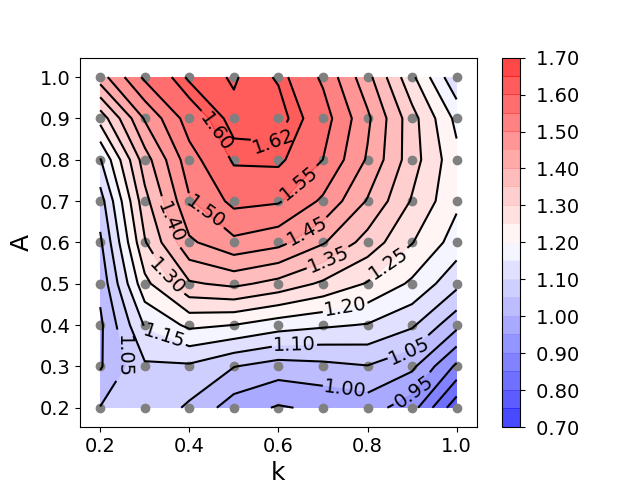}}
    \quad
    \subfigure[\;]{
    \label{Figure19b}
    \includegraphics[width=.45\textwidth]{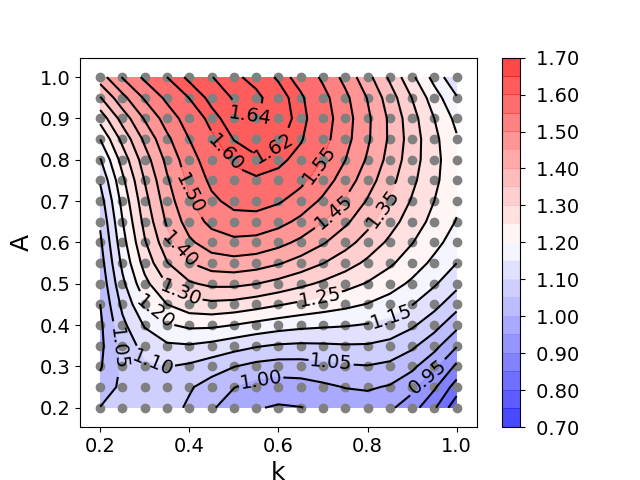}}
    \caption{\label{Figure19} Design points (gray circles) obtained by uniform sampling and corresponding power factor contours. (a) Sparse sampling (${9}\times {9}$ sampled points). (b) Dense sampling (${17}\times {17}$ sampled points).}
\end{figure}

The optimization method also has other advantages explained by the following examples. 
Figure~\ref{Figure19} shows the simulated distribution of the power factor. The power factor values on the scattered points come from the evaluation of the GP model, and these displayed contours are drawn with the hypothesis that the evaluation values are true values. It is seen that simple parameter studies require more sample points (only 35 sample points are used in Figure~\ref{BO_PFm_15}) to obtain comparative information about the power factor distribution and to find the approximate optimal value. This highlights the efficiency of exploring the design space based on the present optimization method.

\section{\label{sec4}Conclusion}
We optimized the kinematics to a gliding flat-plate with spanwise oscillation to enhance the power efficiency by utilizing the portfolio allocation framework of Bayesian optimization, in which the GP model and the hybrid acquisition strategy are adopted. 
The hybrid acquisition strategy improves the robustness of the method with three types of acquisition functions and three different balance parameters for each type of acquisition function.  
The integrated framework is first validated by the analytical cosine mixture function and then applied to design the spanwise oscillation of a gliding flat-plate with the objective of maximizing the power factor. 
This results capture the optimal power factor of 1.65 generated at the non-dimensional oscillating amplitude of 1.00 and the reduced frequency of 0.49. 
The distribution of the power factor indicates that there is an optimal reduced frequency for the power efficiency at the oscillating amplitudes above 0.40. From another perspective, increasing the amplitude has the potential to obtain the higher power efficiency. The underlying physics needs to be further investigated for understanding the high-power-efficiency mechanism.

\begin{acknowledgments}
This work is supported by the National Natural Science Foundation of China (Nos. 11922214 and 91752118), and the National Numerical Windtunnel project. The computations are conducted on Tianhe-1 at the National Supercomputer Center in Tianjin.
\end{acknowledgments}

\section*{Data Availability Statement}
The data that support the findings of this study are available from the corresponding author
upon reasonable request.

\nocite{*}
%

\end{document}